\documentclass[draft]{agujournal2019}
\usepackage{url} 
\usepackage{lineno}
\usepackage[inline]{trackchanges} 
\usepackage{soul}
\draftfalse

\journalname{XXX}

\begin{document}

%
%

\title{Adaptive degenerate space method for source term estimation using a backward Lagrangian stochastic model}

\authors{O.Buchman\affil{1}, E.Fattal\affil{1}}

\affiliation{1}{Applied Mathematics Department,\\Israel Institute for Biological Research, Ness Ziona, Israel}






\correspondingauthor{Omri Buchman}{omrib@iibr.gov.il}
\correspondingauthor{Eyal Fattal}{eyalf@iibr.gov.il}




\begin{keypoints}
\item Source term estimation
\item Backward Lagrangian stochastic model
\item Optimal design
\end{keypoints}

\begin{abstract}
\justify
The general problem of characterizing gas source parameters based on concentration measurements is known to be a difficult task. As many inverse problems, one of the main obstacles for accurate estimation is the non-uniqueness of solution, induced by the lack of sufficient information. As the number of detectors is lowered, which is more than a plausible scenario for many practical situations, the number of possible solutions that can characterize the source increases dramatically, leading to severe errors. In this paper, a Lagrangian stochastic based method for identifying these suspected points, which will be referred to as 'degenerate space', is formulated and analyzed. Then, a new procedure for quantitative prediction of the effect of deploying a new detector in space is used to design an adaptive scheme for source term estimation. This scheme has been tested for several scenarios and is shown to reduce dramatically the degeneracy formed by insufficient information. The combined formulation of degenerate space with the new adaptive scheme is shown to give an accurate estimation of the source parameters for small number of detectors. 
\end{abstract}

\section{Introduction}
\justify

Solving the source estimation problem, i.e. identifying the source position and emission rate based on a given set of measurements, has great importance for many practical situations. For example, estimating the source parameters is a crucial preliminary step for applying a dispersion model in the process of risk assessment of biological or chemical release, as well as for identifying the source in accidental releases in industrial areas.

In the past twenty years a significant progress has been made in the development and testing of new methodologies for solving the source estimation (STE) problem. A detailed review on possible methodologies for solving this problem is beyond the scope of this paper, and can be found in other places \cite{Hutchinson2017}. Briefly, the two most common approaches for solving the STE problem are the optimization or probabilistic approaches. In the optimization approach, one seeks the optimal combination of the source parameters that minimizes a cost function (usually taken as sum of the squared differences between predicted and observed concentrations). In contrast, the objective of the probabilistic approach is the construction of a probability density function of the source parameters based on Bayesian inference theory \cite{{Yee2008},{Yee2012}}. 

Regardless of the chosen approach, a reliable dispersion model is mandatory for retrieving the source parameters. This imposes a computation challenge since the number of dispersion model operations along the STE process may be too demanding for practical use. A significant reduction of the computational effort without loss of accuracy can be achieved by implementing the adjoint source-receptor relationship rather then the standard (forward) dispersion model as done by several authors \cite{Issartel2003, Issartel2005,Keats2007,keats2009,Kumar2015}, and will be described later for the Lagrangian stochastic dispersion model used in this study. 

One of the main difficulties in solving this problem is that as in many inverse modeling problems, the determination of source parameters from measurements is known to be an ill-posed problem, i.e the solution may not exist, may not be unique or suffers from discontinuity. Clearly, adding more information by increasing the number of detectors will gradually solve this problem, but the deployment of a dense array of detectors for covering a wide area may not be feasible. For example, about 40 detectors were used for covering area of less then $200m \times 200m $ in the the well-known Mock Urban Setting Trial (\cite {Yee2004}) which is frequently used for testing STE methods. Trying to maintain the density of detectors used in a wider area may be very demanding for many applications. 

A possible solution for this obstacle can be found by using mobile detectors rather than a static network of detectors, which allows to spare the need of pre-deployment of a dense array of detectors. More than that, the mobility of the detectors enables the possibility to update the position of the detectors along the event in order to maximize the expected information gained by the measurements. Since the general optimization problem of designing a network of detectors by individually placing a set of detectors over a finite grid of points is NP-hard \cite{Keats2010}, the extension of an existing detector network is a more approachable problem. Therefore a proposal mechanism for choosing the position of the next measurement based on given set of already measured concentrations is the crucial step in such methods.  

Several strategies for such dynamic deployment mechanism has been proposed. for example,  n the 'Infotaxis' approach \cite{Vergassola2007}, the searcher chooses the move that maximizes the expected reduction in entropy of the posterior probability field, which amounts to having less uncertainty on the source parameters. In the 'Entrotaxis' approach \cite{Hutchinson2018}, based on the Maximum Entropy sampling principles \cite{Sebastiani2000}, the entropy of the predictive measurement distribution is taken as the reward function, which guides the searcher to where there is the most uncertainty in the next measurement outcome. Ristic  \cite{Ristic2016a} compered number of search strategies based on different information rewards functions for determining the location of a diffusive source in turbulent flows. Keats\cite{Keats2010} applied the Bayesian adaptive exploration (BAE) methodology\cite{Lindley1956,Loredo2004}, which provides a general methodology for choosing how future experiments should be performed so that information about the phenomenon of interest is maximized, to add a new detector to an existing array of detectors deployed according to the original layout in the prery grass project. \hfill \break

In this work we take a different approach for choosing the position of the new detector.
As in many applications, after discretizing the parameter space, the STE problem can be written as a linear inverse problem \cite{Menke}. Yet, although its linearity solving this problem imposes a great challenge since in practice the small number of available measurement is transformed into a highly under-determined problem. The objective of this paper aimed directly to reduce this non uniqueness by two steps: first a simple and efficient mechanism will be  proposed for identifying the suspected points in the parameter space based on a given set of measurements. The method for the construction of this sub-space, which will be refereed a "degenerate space", will be described in section \ref{construction_of_degeneracy}. Then a criterion for choosing the location for a new detector will be presented, based on a statistical evaluation of the expected degeneracy reduction as function of the detector's position. These two steps can be combined iteratively to generate an adaptive algorithm which exploits the given information from a set of “old” measurements to plan the next measurement. Similar to \cite{Keats2010}, we examine the performance of this method of a quasi- steady-state release for several scenarios differing by the initial locations of the detectors. The algorithm is shown to converge fast, sparing the need for a large number of pre-deployed detectors, as will be described in section \ref{testing_adaptive_method}.

\section{Theory}
\subsection{The Lagrangian stochastic model}
In order to solve the source estimation problem, one has to specify the underlying turbulent pollutant dispersion model. In this work, a Lagrangian stochastic model (LSM) developed at IIBR \cite{Fattal2014r,Fattal2021,Fattal2023}, has been used. This modelling approach is known to be able to describe consistently gas dispersion phenomena, in rather complicated atmospheric scenarios such as non-homogeneous turbulent regimes, complex terrain and canopies (urban and vegetation), and is known to be superior to advection-diffusion based approaches (see e.g. \cite{Gavze2018}). The LSM has been described in detail in many articles \cite{Thomson1987,Flesch1995}. Briefly, the basic idea is to propagate the position and velocity of the Lagrangian fluid particles according to the Langevine equations:
\begin{linenomath*}
\begin{equation}
dx_i=u_idt
\end{equation}
\end{linenomath*}
\begin{linenomath*}
\begin{equation}
du_i=a_i(x,u,t)dt+b_{ij}(x,u,dt)dW_j(t)
\end{equation}
\end{linenomath*}
where $x$ and $u$ are the position and velocity of the particles and $dW_i$ is a random increment selected from a Normal Distribution with variance $dt$. The deterministic coefficients $a_i$ can be determined via the Fokker-Planck equation, by satisfying the well mixed (necessary) condition \cite{Thomson1987}. According to the Kolmogorov similarity theory \cite{pope_2000}, the stochastic term  takes the form of $b_{ij}=\sqrt{(C_0 \epsilon)} \delta_{ij} $ where $\epsilon$ is the average dissipation rate of the kinetic energy.

By counting all ‘touchdowns’ of the stochastic particles in some volume which represents the detector, one can calculate the probability density $P^f(r,t|r',t')$ that a particle spreads from $r'$ at time $t'$  will reach the detector located at point $r$ after time $t$. Flecsh and Wilson \cite{Flesch1995} showed that by modifying the Langevine equation, one can propagate the particles backwards in time (BLSM), i.e. from the future to the past, and that for an incompressible fluid the following relation holds:
\begin{linenomath*}
\begin{equation}
P^f(r,t|r',t' )=P^b (r',t'|r,t)
\end{equation}
\end{linenomath*}
where $P^b(r',t'|r,t)$ is the probability density function that a particle evolving from the detector located at $r$ and propagated backwards in time, will reach the 'source' located at point $r'$ at time $t'$.
The transition probability for such a process can be related to the ensemble averaged concentration \cite{Flesch1995,Sawford1985} by:
\begin{linenomath*}
\begin{equation}
d(r,t)=\int\int\limits_{0}^{\infty} S(r',t')P^f (r,t|r',t')dt'dr'
\end{equation}
\end{linenomath*}
where $S(r',t')$ describes the spatial and temporal source distribution in $kg \times m^{-3} \times sec^{-1}$. 
The calculated concentration, averaged over the detector's volume $v_d$ which is centered at position $r$, at stationary turbulence and sustained uniformly distributed source can be written as:
\begin{linenomath*}
\begin{equation}\label{eq:forward_backward_relation1}
d(r)=\frac{q}{V_d}\int\limits_{v_{d} }\left( \int\limits_{v_s } \int\limits_{0}^{\infty} P^f (r,t|r',0 )dtdr' \right)dr
\end{equation}
\end{linenomath*}
where q is the emission rate, $v_d$ is the volume of the detector and $v_s$ is the volume of the source.

As will be described in the next section, the parameter space $\Theta$ will be discretized by setting a grid of $N_x\times N_y\times N_q$ cells. Each hypothesis about the source will be labeled as $\theta_j=(r_j,q_j )\in \Theta $, where $r_j=(x_j,y_j )$ are the coordinates of the center of the j’th cell and $q_j$ is the emission rate. Therefore the concentration at the i’th detector, located $R_i$ due to a source characterized by the j’th hypothesis is given by:
\begin{linenomath*}
\begin{equation}\label{eq:d_ri_theta_k}
d(R_i|\theta_j)=\frac{q_j}{V}\int\limits_{v_{i} }\left( \int\limits_{v_j} \int\limits_{0}^{\infty} P^f (r,t|r',0 )dtdr' \right)dr=\frac{q_j}{V}\int\limits_{v_{j} }\left( \int\limits_{v_i} \int\limits_{0}^{\infty} P^b (r',0|r,t )dtdr \right)dr'
\end{equation}
\end{linenomath*}
where $v_j$  is the j’th cell in the parameter space and $v_i$ is a small volume centered on $R_i$. The last passage in \ref{eq:d_ri_theta_k}, due to the Forward-Backward relation, can be very useful, as will be seen later.

\subsection{The degenerate space}\label{construction_of_degeneracy}
Assume that there is a discrete hypotheses space $\Theta$, and for every hypothesis $\theta \in \Theta$ the expected concentration at some point in space $d(\theta)$ can be calculated. The source is fully characterized by one and only one hypothesis $\theta_{ex} \in \Theta $. The  model-measurement deviation $\sigma$ represents the maximal discrepancy between the measured concentration $d$ and the calculated concentration if the source is characterized by the exact hypothesis, i.e :
\begin{linenomath*}
\begin{equation}\label{eq:nodel_measurment_discrapncy}
d\in[d(\theta_{ex} )-\sigma,d(\theta_{ex} )+\sigma]
\end{equation}
\end{linenomath*}
In this work we shall assume that the error has linear dependency on the concentration in order to retain the relative error fixed. In order to avoid unrealistic small errors associated with low concentrations, a constant term $\sigma_1$ will also be added, so the error will be: 
\begin{linenomath*}
\begin{equation}
\sigma(\theta)=\sqrt{(\kappa d(\theta))^2+\sigma_1^2}
\end{equation}
\end{linenomath*}
where $\kappa$ is some proportional constant. 

Now assume that we have measured concentration $d$. Which hypotheses are important in this case? Consider an arbitrary hypothesis $\theta_k$. If it was the exact hypothesis, the measurement outcome should be in the range described by equation (\ref{eq:nodel_measurment_discrapncy}). Formally:
\begin{linenomath*}
\begin{equation}
\theta_k=\theta_{ex} \Rightarrow  d\in \left[ d(\theta_k) - \sigma(\theta_k),d(\theta_k )+\sigma(\theta_k ) \right].
\end{equation}
\end{linenomath*}
Note that there might be other hypotheses for which the R.H.S holds. The negation of the last claim is more useful:
\begin{linenomath*}
\begin{equation}
d\notin[d(\theta_k )-\sigma(\theta_k ),d(\theta_k )+\sigma(\theta_k )]\Rightarrow\theta_k \not = \theta_{ex}
\end{equation}
\end{linenomath*}
This is a good way to select the good hypotheses from the bad ones - for a given measurement, eliminate all the hypotheses for which $ d\notin \left[ d(\theta_k )-\sigma(\theta_k ),d(\theta_k )+\sigma(\theta_k )\right ]$ because they cannot be considered as the exact hypothesis. 
This is the basis for the “Degenerate space”, defined as the set of all equally probable hypotheses for the source. If there are M measurements, $d_1\ldots d_M$ located at positions $R_1\ldots R_M$,  define the M-detectors degenerate space $S(R_1 \ldots R_M)$ as:
\begin{linenomath*}
\begin{equation}\label{eq:deg_sapce_def}
S(R_1\ldots R_M) = \{\theta_k \in \Theta \left|d(R_i|\theta_k )-d_i \right| \leq \sigma(\theta_k ),i=1 \ldots M\}
\end{equation}
\end{linenomath*}
where $d(R_i|\theta_k)$ designates the calculated concentration due to k’th hypothesis at the i’th detector’s position, calculated by equation (\ref{eq:d_ri_theta_k}). The last equation can be written recursively as:
\begin{linenomath*}
\begin{equation}\label{eq:recursive_deg_sapce_def}
S(R_1\ldots R_M) = \{ \theta_k \in S(R_1 \ldots R_{M-1}) \left|d(R_M|\theta_k )-d_M \right| \leq \sigma(\theta_k )\}
\end{equation}
\end{linenomath*}
which will be useful later.

%

the construction of $S(R_1...R_M)$ by applying equation (\ref{eq:deg_sapce_def}) imposes a severe computational obstacle since for every hypothesis in the parameters space, the expected concentration at every detector must be calculated, and then only the hypotheses for which the criterion holds are included. The number of required LSM operations for this task is the number of different cells used to discretize $xy$ plane which can be very large in practical situations (for example in the following analyzed cases the number of $xy$ cells is $ 2.5 \times 10^5 $). Since the time of typical LSM simulation is typically about 30 min, the total computational time could take weeks, even by operating on 100 parallel processors. 

A significant reduction of the computational time can be achieved by introducing a backward Lagrangian stochastic model (BLSM), since a single BLSM simulation enables the calculations of the concentration at a specific detector from all the cells in $xy$ plane of the parameter space (see equation (\ref{eq:d_ri_theta_k})). Hence the number of BLSM simulations is the number of detectors rather than the number of $xy$ cells, which is much smaller, leading to a dramatic reduction of the total computational effort. To conclude, the degenerate space can be calculated efficiently by $M$ BLSM simulations by combing its definition and the right part of equation (\ref{eq:d_ri_theta_k}) as $d(R_i|\theta_k)$.

\subsection{Adaptive degeneracy reduction} \label{ADR_sectioin}
Now let’s take another step forward. After the construction of $S(R_1 \ldots R_M)$ we are allowed to perform another measurement in any place we wish. Where should we place the new measurement? Clearly a “good” measurement will reduce dramatically the degenerate space, while a “bad” one leaves the degenerate space almost the same. This intuition can be quantified by small or large (up to 1) value of the reduction factor, defined as $\Gamma(R_{M+1}) =\frac {|S(R_1 \ldots R_{M+1})|} { |S(R_1 \ldots R_M )|}$. Note that the reduction factor can be calculate only after the new detector had been deployed at $R_{M+1}$ since the measurement $d_{M+1}$ is crucial part of the degenerate space definition.

How can we predict the expected reduction as function $R_{M+1}$ of before the deployment of the new detector? If $\theta_j=\theta_{ex}$ and the measurement outcome is $d_{M+1}$  then the new degenerate space will be:
\begin{linenomath*}
\begin{equation}\label{eq:condtioned_S}
S(R_{M+1}|\theta_j) = \{\theta_k \in S(R_1 \ldots R_M)  \left|d(R_{M+1}|\theta_k)-d_{M+1} \right| \leq \sigma(\theta_k )\}
\end{equation}
\end{linenomath*}
and the conditioned reduction factor will be defined as:
\begin{linenomath*}
\begin{equation}\label{eq:conditioned_Gamma}
\Gamma(R_{M+1}|\theta_j)=\frac{|S(R_1 \ldots R_{M+1}|\theta_j)|} {|S(R_1 \ldots R_M )|} 
\end{equation}
\end{linenomath*}
(we omitted the dependency of the reduction factor on the already deployed detectors $(R_1 \ldots R_M)$ since their positions cannot be changed in contrast to the position of the new detector). 
As mentioned before, this is not useful since $d_{M+1}$ can be known only after the deployment and measurement of the new detector at $R_{M+1}$. Therefore define the readily attainable set:
\begin{linenomath*}
\begin{equation}\label{eq:conditioned_s}
s(R_{M+1}|\theta_j) = \{\theta_k \in S(R_1 \ldots R_M)  \left|d(R_{M+1}|\theta_k)-d(R_{M+1}|\theta_j) \right| \leq \sigma(\theta_k)+\sigma(\theta_j)\}
\end{equation}
\end{linenomath*}
And its corresponding conditioned reduction factor: 
\begin{linenomath*}
\begin{equation}\label{eq:conditioned_gamma}
\gamma(R_{M+1}|\theta_j)=\frac{|s(R_1 \ldots R_{M+1}|\theta_j)|} {|S(R_1\ldots R_M )|} 
\end{equation}
\end{linenomath*}
Now let’s relate $\gamma(R_{M+1}|\theta_j)$ to $\Gamma(R_{M+1}|\theta_j)$. By definition for every  $\theta_k\in S(R_{M+1}|\theta_j)$ :
\begin{linenomath*}
\begin{equation}\label{eq:first_assumption1}
\left| d(R_{M+1}|\theta_k)-d_{M+1} \right| \leq \sigma(\theta_k )
\end{equation}
\end{linenomath*}
In addition since we assumed that $\theta_j=\theta_{ex}$ then from (\ref{eq:nodel_measurment_discrapncy}):
\begin{linenomath*}
\begin{equation}
\left| d(R_{M+1}|\theta_j)-d_{M+1} \right| \leq \sigma(\theta_j )
\end{equation}
\end{linenomath*}
The combination of the last two inequalities leads to: 
\begin{linenomath*}
\begin{equation}
|d(R_{M+1}|\theta_k)-d(R_{M+1}|\theta_j)|\leq |d(R_{M+1}|\theta_k )-d_{M+1}| + |d(R_{M+1}|\theta_j )-d_{M+1}| \leq \sigma(\theta_j )+\sigma(\theta_k )
\end{equation}
\end{linenomath*}
So we showed that $\theta_k \in s(R_{M+1}|\theta_j )$ and therefore $S(R_{M+1}|\theta_j )\subseteq s(R_{M+1}|\theta_j)$, which implies that for every $\theta_j$: 
\begin{linenomath*}
\begin{equation}\label{eq:cond_gamma_Gamma_relation}
\Gamma(R_{M+1}|\theta_j )\leq \gamma(R_{M+1}|\theta_j )
\end{equation}
\end{linenomath*}
Now consider the probability that if we choose randomly a hypothesis from $S(R_1 \ldots R_M )$, the reduction factor and its estimator will be smaller than some small number $\alpha$. Since all points are equally probable this can be calculated as:
\begin{linenomath*}
\begin{equation}
Pr(\Gamma(R_{M+1})\leq\alpha)=\frac{\left|\{\theta_i| \Gamma(R_{M+1}|\theta_i)\leq \alpha \}\right|}{|S(R_1 \ldots R_M)|}
\end{equation}
\end{linenomath*}
And its “companion”: 
\begin{linenomath*}
\begin{equation}
Pr(\gamma(R_{M+1})\leq\alpha)=\frac{\left|\{\theta_i| \gamma(R_{M+1}|\theta_i)\leq \alpha \}\right|}{|S(R_1 \ldots R_M)|}
\end{equation}
\end{linenomath*}
From equation (\ref{eq:cond_gamma_Gamma_relation}) we know that for every $\theta_j$ for which $\gamma(R_{M+1}|\theta_j )\leq \alpha$ we get $\Gamma(R_{M+1}|\theta_j )\leq\gamma(R_{M+1}|\theta_j )\leq\alpha$ so $\{\theta_j|\Gamma(R_{M+1}|\theta_j )\leq\alpha \} \subseteq \{\theta_j|\gamma(R_{M+1}|\theta_j )\leq\alpha\}$ and we showed that:
\begin{linenomath*}
\begin{equation}\label{eq:_Pr_gamma_Pr_Gamma_relation}
Pr(\Gamma(R_{M+1} )\leq \alpha) \geq Pr(\gamma (R_{M+1}) \leq \alpha)
\end{equation}
\end{linenomath*}
Clarification of the last claim can be seen by the following example. Assume that we have 100 points and we obtain $Pr(\gamma\leq 0.2)=0.6$. Therefore, there are $60$ points for which $\gamma(R_{M+1}|\theta_j )\leq 0.2$ and 40 points for which $\gamma(R_{M+1}|\theta_j )\geq 0.2$ . For each point from the first class we know $\Gamma(R_{M+1}|\theta_j )\leq \gamma(R_{M+1}|\theta_j )\leq0.2$ and therefore $ Pr(|\Gamma(R_{M+1} )\leq 0.2)\geq 0.6$.

Recall that our objective was to estimate before deployment the expected reduction of the degenerate space as function of the new detector position. While direct estimation of this quantity is not accessible, we can bound it from below by the attainable quantity $\gamma$ as shown in equation (\ref{eq:_Pr_gamma_Pr_Gamma_relation}). One possible strategy that can be taken from here is to search for the candidate $R^*_{M+1}$  which maximizes of $Pr(\gamma(R_{M+1} )\leq \alpha)$. Although this choice does not necessarily maximizes $Pr(\Gamma(R_{M+1})\leq\alpha)$, we are guaranteed that:
\begin{linenomath*}
\begin{equation}\label{eq:Fundamental_eqaution}
Pr(\Gamma(R^*_{M+1} )\leq\alpha) \geq \max_{R_{M+1}}Pr(\gamma(R_{M+1} )\leq\alpha)
\end{equation}
\end{linenomath*}

What is the meaning of this criterion? equation \ref{eq:Fundamental_eqaution} ensures that the probability that the reduction factor is smaller than a chosen $\alpha$ is larger than the optimized value. For the rest of the paper, we shall use $\alpha=0.2$, meaning that the reduction of degenerate space is at least by factor 5. As we shall see later, in many cases the probability for such a reduction will be larger than 0.6.

Equation (\ref{eq:Fundamental_eqaution}) can be used iteratively, where in every cycle the the degeneracy of the previous step is used to deploy the new detector. A schematic description for such algorithm is:
\hfill \break
while $N_d>toll$ :
\begin{enumerate}
  \item Prediction: based on $S(R_1\ldots R_M)$, use equation (\ref{eq:Fundamental_eqaution}) to estimate $R_{M+1}^*$
  \item Measurement: deploy the new detector at $R_{M+1}^*$ and measure $d_{M+1}$
  \item Reduction: use equation (\ref{eq:recursive_deg_sapce_def}) to construct the new degenerate space  $S(R_1\ldots R_M,R_{M+1}^*)$
  \item Calculate $N_d=|S(R_1 \ldots R_M,R_{M+1}^*)|$ and return to step 1
\end{enumerate}
where $toll$ and $\alpha$ have to be specified by the user and $N_d$ is the number of degenerate points.

\section{Results}
\subsection{Numerical details} \label{Numerical_details}
Since our main interest in this section is the study of the proposed mechanism for deployment a new detector based on previous measurements, we shall focus on the relatively simple meteorological scenario and use a the same dispersion model used for the source estimation to generate the input concentrations ('synthetic data'). In all examined cases the external wind is 2 m/sec in the east direction in neutral stability. Unless mentioned otherwise, the source is located at the center of $10km\times10km$ grid composed by cells of $20m\times20m$ and the emission rate of the source is $1 kg/min$. 
The Lagrangian stochastic model is used to calculate the concentration based on 400000 Lagrangian particles.

As mentioned in section \ref{ADR_sectioin}, the construction of the degenerate space at each cycle requires the specification of the error  dependency on the concentration. In order to estimate this relation a serious of 30 identical LSM operations was performed, differing by their random seeds. Based on these calculations, the ratio between the standard deviation and the average concentration was calculated at 7 typical points located downstream to the source position. This ratio, averaged over these points, is $0.15 \pm 0.14$, which implies that the proportional constant should be $\kappa=0.3$. In addition, the fixed term in the error expression (see equation \ref{eq:nodel_measurment_discrapncy}) was taken as $\sigma_1=1E-8$. 

\subsection{Testing the adaptive source algorithm}\label{testing_adaptive_method}
\begin{figure}[h!]
\centering
\noindent\includegraphics[width=100mm]{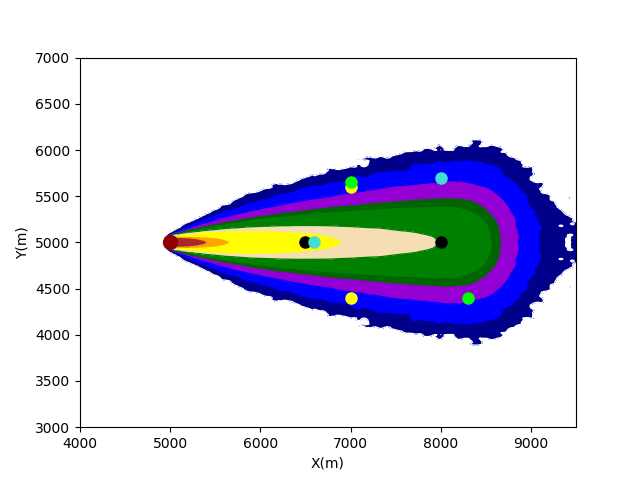}
\caption{The concentration map, generated by a source located at $(5000m\times5000m)$ is shown. The position of the two detectors for the first four cases are also shown, each case represented by a different color (black for case 1, yellow for case 2, green for case 3 and cyan for case 4)}
\label{fig:fig_2}
\end{figure}

We shall examine the adaptive source term estimation algorithm, specified in section \ref{ADR_sectioin}, in six cases differing by the initial arrangement of the first two pre-deployed detectors and the source parameters. The concentration map generated by operating the Lagrangian stochastic model for source located at $(5000,5000)$, is shown in Figure \ref{fig:fig_2}. A different color is used to designate the pair of detectors of the first four cases. In the first case (black points) the first two detectors are pre-deployed along the same line parallel to the wind direction. In the second case (yellow points) the detector aligned the same line perpendicular to the external line. In cases 3 (green) and 4 (cyan) the symmetry was removed. The fifth case is similar to the forth case but the source location has been shifted to (4000,5200). In the sixth case the setup is identical to the third case, but the emission rate is 2 kg/min. 

\begin{figure}[h!]
\centering
\includegraphics[width=.85\linewidth]{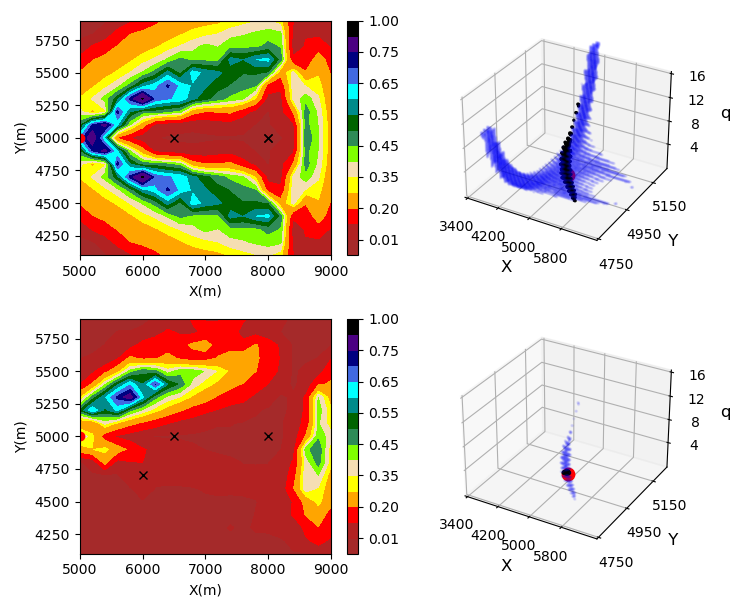}
\caption{Case 1 analysis: Each row describes a cycle of the algorithm, starting from two detectors located at (6500,5000) and (8000,5000). In the left side of each row, the expected reduction according to the criterion described in equation (\ref{eq:Fundamental_eqaution})  is shown for all possible locations of the new detector. The already deployed detectors, used for the construction of the degenerate space in the beginning of the cycle, are designated by black crosses. The degenerate space before (blue) and after (black) the deployment of the new detector can be seen in the right panel of every row. the actual reduction gained by the procedure can be seen by the difference between the two colors. After two cycles, the degenerate space is reduced almost entirely to the correct source parameters, represented by the red dot.}
\label{fig:fig_3_Case_1}
\end{figure}

\begin{figure}[h!]
  \centering
  \includegraphics[width=.85\linewidth]{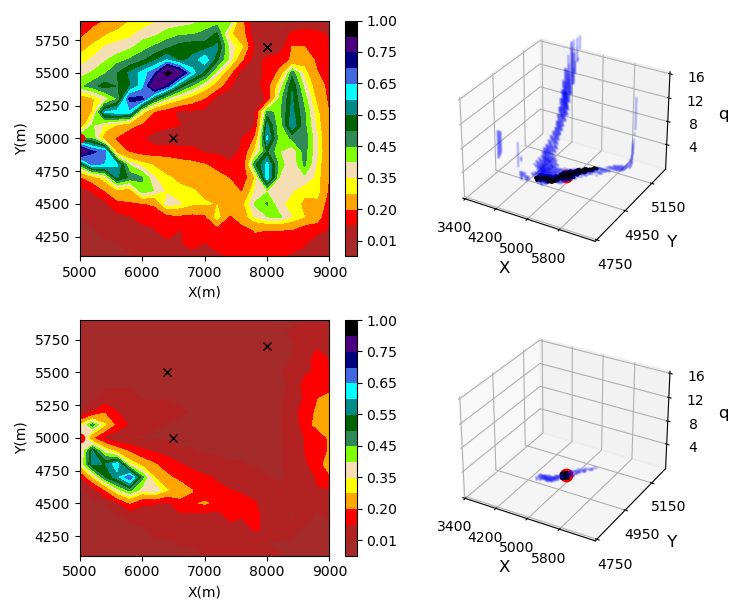}
  \caption{Case 4 analysis: the same analysis as described in the previous figure, starting from two detectors located at (6500,5000) and (8000,5700)}
  \label{fig:fig_4_Case_2}
\end{figure}

In all cases the input concentration for the source estimation process was taken as the value of the map at the detector’s positions. Full analysis cases 1 and 4 can be seen in Figures \ref{fig:fig_3_Case_1} and \ref{fig:fig_4_Case_2} and shall now be described. \hfill \break

\textbf{Case 1}: In the upper-right panel of Figure \ref{fig:fig_3_Case_1}, a map of $Pr(\gamma(R_3)\leq 0.2)$ is shown based on the available measurements at the first two detectors located at $R_1=(6500,5000), R_2=(8000,5000)$ (also shown as black crosses on the map). One can easily see two stretches of preferred points and a wide un-preferred regime enclosed by these two stretches. The position of the third detector was chosen at (6000,4700) for which the criterion is maximized. The effect of this choice on the degeneracy reduction can be seen in the upper right panel of Figure \ref{fig:fig_3_Case_1}. The degenerate space before and after the third measurement can be seen in blue and black points correspondingly. Note that the wide and spread original space that contained $8704$ points, has been reduced to a narrow diagonal stretch containing only $460$ points. \hfill \break
The first row in figure \ref{fig:fig_3_Case_1} describes the first cycle in the operation of the adaptive algorithm for case 1. The next row can be understood in a similar way – in the left panel a map of the predicted reduction is formed based on the degenerate space of the previous iteration (note that a new black cross was added at the optimized position of the previous cycle), and a point which optimizes this map is chosen as the position of the new (fourth) detector. The effect of the new measurement can be shown in the right panel in the second row. As before, blue points represent the degenerate space before the addition of the new detector, which is the outcome of the last cycle, and black points represent the remaining points after the addition of the new information. \hfill \break
Note that the initial arrangement of the detectors along a line parallel to the wind direction, led to a very large and spread initial degenerate space. Nevertheless, every addition of detector reduces dramatically the size and shape of the degenerate space of the last cycle, so after deployment of two new detectors, the degenerate space is condensed to the small proximity of the source, represented by the red dot in the lower right panel. \hfill \break
The performance of the adaptive algorithm along two cycles for the first case are summarized in the upper part of table \ref{table:performance_6_cases}. For each cycle the number of degenerate points, as well as the averaged source parameters and their corresponding standard deviations are shown (the actual source position and emission rate were subtracted from the averages). One can clearly see fast convergence in all these criteria to the correct value.\hfill \break

\textbf{Case 4}: In Figure \ref{fig:fig_4_Case_2}, a similar analysis for the fourth case (cyan dots in Figure \ref{fig:fig_2}) is shown. The asymmetric deployment of the first two detectors dictates an asymmetric shape of the probability map. In addition, the regime of high probability is far more narrow and difficult to predict intuitively without a detailed calculation. In this case, the initial averages and standard deviations of all parameters are large, as can be seen in the first row of case 4 in table \ref{table:performance_6_cases}. Adding new detectors dramatically reduces the error of all parameters, as can be seen in the following rows of the table, as well as by comparing the blue and black points in the figure.

A summary of all six cases is shown in table \ref{table:performance_6_cases}. The number of cycles is limited by the demand that the number of degenerate points is less than some small number (taken as 100 points in this work). One can see that in all cases, a rapid decrease in the number of suspected points occurs after the first iteration (i.e. by adding the third detector). Furthermore, both the averaged parameters (after subtraction of the exact values) and the standard deviation of the degenerate space reduces along the procedure (the errors in y direction is typically much smaller then the x direction because the wind is aligned along the x axis).
Note that the search was done in parameter space of $10000m\times10000m$, so the relative error of every spatial parameter, normalized by the corresponding length scale, at the end of the algorithm operation is very small. More than that, the resolution of the source estimation is limited by the size of the grid cells used to predict the concentration from the LSM operation, which was $20m\times20m$ for the spatial coordinates and $0.15$ for the emission rate in these cases. Therefore, in all cases, the errors are within deviation of three spacial cells at the most.
Note that since the wind direction is aligned along the x axis, the errors in this direction (both the average and standard deviation) is much larger then along the y axis. 

\begin{table}[h]
\caption{The results of the adaptive source estimation algorithm for six cases are shown in the table. For every case, the first row describes the features of degenerate space before applying the adaptive algorithm, based on two pre-deployed detectors. At every row, the number of degenerate points, the averages of the source parameters (after subtraction of the exact value) and their corresponding standard deviations are shown}
\centering
\begin{tabular}{c c | c| c c c| c c c}
\hline
cycle & $N_{detectors} $ & $N_d$ & $\langle x \rangle-x_{sor}  $ & $\langle y \rangle-y_{sor}  $ & $\langle q \rangle -q_{sor} $ & $\sigma_x$ &	$\sigma_y$ & $\sigma_q$  \\
\hline
CASE 1  & & & & & & & \\
  & 2 &	8704 & 231 & 0 & 3.76 & 400 & 210 & 4.24 \\
1 &	3 &	460 & 130 & 32 & 0.68 & 296 & 72 & 0.92 \\
2 &	4 & 32 & 32 & 0 & 0.12 &	45 & 0 & 0.19 \\
\hline
CASE 2 & & & & & & & \\
  &	2 & 2583 & 896 & 0 &3.53 &	1062 & 22 &	3\\
1 &	3 &	235 &	87 & 0.6 & 0.75 &	394 &	11 & 1.2\\
2 &	4 &	12 & 45 & 0 & 0.18 & 27 & 0.0 &	0.12\\
\hline
CASE 3 & & & & & & & \\
  &	2 &	2183 & 30 & 30 & 3.9 & 630 & 139 & 3.1\\
1 &	3 &	73 & 39 & 11 & 0.21 &	74 & 12 &	0.25\\
2&	4&	31 &	30&	12&	0.06&	40&	12&	0.15\\
\hline
CASE 4 & & & & & & & \\
 & 2 & 3762 &	451 & 87 & 3.5 &	604 &	171 &	3.9\\
1 & 3 &	441 &	56 & 24 & 0.35 &	186 &	53 & 0.45\\
2 &	4 &	43 & 13 & 5 & 0.12 & 60 & 8 &	0.18\\
\hline
CASE 5 & & & & & & & \\
 &	2 &	2743 & 428 & 2 & 3.5 & 771 & 102 &3.22\\
1 &	3 &	245 &	120 & 4 & 0.21 &	188 &	19 &0.25 \\
2 &	4 & 24 & 70 &	5 & 0.08 & 94 & 13 & 0.14 \\
\hline
CASE 6 & & & & & & & \\
 &	2 &	3060 &	2 & 50 & 5.3 &	575 &	117 &	3.3\\
1 &	3 &	158 & 49 &	10 & 0.46 & 78 & 12 & 0.49 \\
2 &	4 &	63 & 28.5 & 11 &	0.19 & 39 & 12 & 0.32 \\
\hline
\end{tabular}
\label{table:performance_6_cases}
\end{table}

\section{Summary and Discussion}
In this paper we have presented a new approach for an adaptive deployment of detectors, aimed to reduce the existence of many hypotheses with approximately the same probability for describing the source parameters. This adaptive approach relies on two concepts – the identification of the suspected points based on current measurements, and the prediction the effect of the new detector's position. 

The construction of the degenerate space defined in section \ref{construction_of_degeneracy} can be intuitively thought as the intersection of the detectors "iso-surfaces", which are the sets of all points in the parameter space that will retrieve the actually measured concentration in every detector. This idea is somewhat similar in its character to the approach presented by Keats \cite{Keats2007}, who demonstrated how the ‘regimes of influences’, generated by solving the adjoint diffusion-advection equation for every detector, can be used to select the possible locations of the source. There is however a significant difference between the two approaches since the construction of these regimes does not relies on the concentrations that were actually measured, in contrast to the definition of the degenerate space given before.

\begin{figure}[h!]
\centering
\noindent\includegraphics[width=120mm]{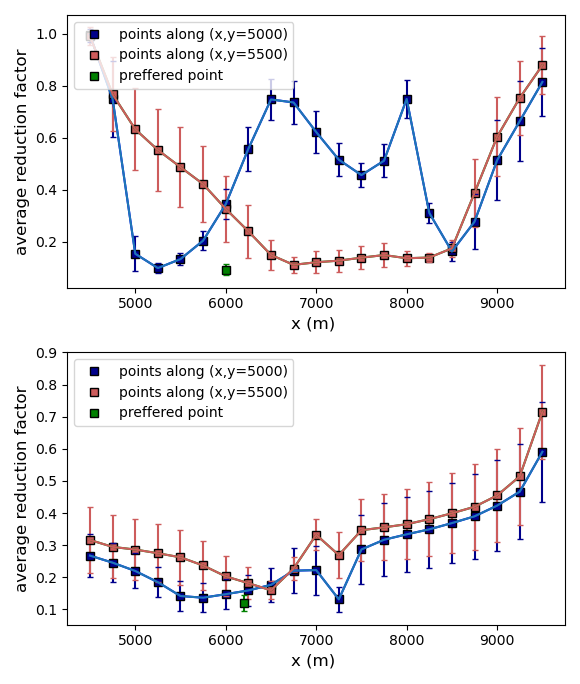}
\caption{The actual reduction, averaged over all possible hypotheses in the first cycle of the first and second cases, is shown for different points along the line (x,5000) and (x,5500). Note that the best performance is achieved for the preferred point (green point).}
\label{fig:fig_5}
\end{figure}

How should the adaptive method developed in this study be tested? The straightforward answer may be to calculate the actual reduction by deploying the new detector at the proposed location, measure the concentration and calculate the new degenerate space formed by this deployment. Then, this outcome could be compared with the actual reduction formed by the deployment of the new detector at other places. Note however that all the points in the degenerate space can serve as the source and each one of them will reproduce a similar degenerate space. If the starting point would be a different point taken from the degenerate space, applying the previously mentioned test may lead to a completely different reduction. Therefore a better test for the performance of the proposal mechanism is to compare the actual obtained reduction as function of the detectors position, averaged over all equally probable hypotheses.

In the upper panel of Figure \ref{fig:fig_5}, the averaged actual reduction is shown for different possible positions for the third detector in the first case along two lines $(x, 5000)$ and $(x, 5500)$ (error bar was added to indicate the standard deviation associated with the average at every point). The average reduction for the preferred point, chosen by the criterion mentioned in section \ref{ADR_sectioin}, is also shown as green rectangular. In the lower panel the same analysis is shown for the second case. One can see the best outcome is obtained for the proposed position in both cases. In the first case the difference between the preferred and other points can be dramatic for most of the examined points (note that the averaged reduction for the preferred point is less then 0.1, i.e. on average less the 10\% of the degenerate points survived the elimination induced by the third measurement). In the second case the difference is less dramatic, although the reduction for the preferred point is still significantly lower then all other points(0.12).  

In addition, we compared the reduction obtained by deploying the third detector at the proposed position to the reduction obtained by randomly choosing its position. The reduction averaged over 200 random choices taken from a box of $[3000,10000]\times [4000,6000]$ is $0.6$, much larger than the obtain reduction for the proposal position ($0.05$).  

The effect of detector’s number and their arrangement on source estimation accuracy had been addressed by \cite {Rudd2012}, who used the results from wind tunnel experiment as input for gradient based source estimation using Gaussian dispersion model. In their work, 11 different sets of 4 detector's configurations were used to estimate the source parameters in a similar setup as taken here (homogeneous external wind is 1.3 m/sec). As can be seen in the first row of Table 1 in Rudd's paper, the error in the source position divided by a typical length scale, taken as the averaged distance between the source and detectors, is about 0.19 and 0.24 for 4 and 5 detectors correspondingly. In order to compare our results we normalized the error of the second cycle (after 4 detectors had been deployed) with the averaged distance ($\sim  2000m$), yielding almost an order of magnitude smaller error in all 6 cases.

To conclude, in practical situations where the available measurements is expected to small relative to the hypothesis space, the location of the detectors is crucial for retrieving the source parameters correctly. The suggested proposal mechanism seems to predict correctly the effect of degeneracy reduction induced by these measurement, and can be used as for solving the STE problem efficiently.

\acknowledgments
Data were not used, nor created for this research.

%
%

\bibliography{library.bib}

\begin{thebibliography}{}

\bibitem [\protect \citeauthoryear {%
Fattal%
}{%
Fattal%
}{%
{\protect \APACyear {2014}}%
}]{%
Fattal2014r}
\APACinsertmetastar {%
Fattal2014r}%
\begin{APACrefauthors}%
Fattal, E.%
\end{APACrefauthors}%
\unskip\
\newblock
\APACrefYearMonthDay{2014}{}{}.
\newblock
\APACrefbtitle {A non-homogeneous non-Gaussian Lagrangian-stochastic model for
  pollutant dispersion in complex terrain, and its comparison to Haifa 2009
  tracer campaign} {A non-homogeneous non-gaussian lagrangian-stochastic model
  for pollutant dispersion in complex terrain, and its comparison to haifa 2009
  tracer campaign}\ \APACbVolEdTR {}{IIBR Scientific Report,\ \BNUMS\
  2014/56/53/5614, Oct. 2014}.
\PrintBackRefs{\CurrentBib}

\bibitem [\protect \citeauthoryear {%
Fattal%
, David-Saroussi%
, Buchman%
, Tas%
\BCBL {}\ \BBA {} Klausner%
}{%
Fattal%
\ \protect \BOthers {.}}{%
{\protect \APACyear {2023}}%
}]{%
Fattal2023}
\APACinsertmetastar {%
Fattal2023}%
\begin{APACrefauthors}%
Fattal, E.%
, David-Saroussi, H.%
, Buchman, O.%
, Tas, E.%
\BCBL {}\ \BBA {} Klausner, Z.%
\end{APACrefauthors}%
\unskip\
\newblock
\APACrefYearMonthDay{2023}{}{}.
\newblock
{\BBOQ}\APACrefatitle {{Heterogenous Canopy in a Lagrangian-Stochastic
  Dispersion Model for Particulate Matter from Multiple Sources over the Haifa
  Bay Area}} {{Heterogenous Canopy in a Lagrangian-Stochastic Dispersion Model
  for Particulate Matter from Multiple Sources over the Haifa Bay
  Area}}.{\BBCQ}
\newblock
\APACjournalVolNumPages{Atmosphere}{14}{1}{}.
\newblock
\begin{APACrefDOI} \doi{10.3390/atmos14010144} \end{APACrefDOI}
\PrintBackRefs{\CurrentBib}

\bibitem [\protect \citeauthoryear {%
Fattal%
, David-Saroussi%
, Klausner%
\BCBL {}\ \BBA {} Buchman%
}{%
Fattal%
\ \protect \BOthers {.}}{%
{\protect \APACyear {2021}}%
}]{%
Fattal2021}
\APACinsertmetastar {%
Fattal2021}%
\begin{APACrefauthors}%
Fattal, E.%
, David-Saroussi, H.%
, Klausner, Z.%
\BCBL {}\ \BBA {} Buchman, O.%
\end{APACrefauthors}%
\unskip\
\newblock
\APACrefYearMonthDay{2021}{}{}.
\newblock
{\BBOQ}\APACrefatitle {{An urban lagrangian stochastic dispersion model for
  simulating traffic particulate-matter concentration fields}} {{An urban
  lagrangian stochastic dispersion model for simulating traffic
  particulate-matter concentration fields}}.{\BBCQ}
\newblock
\APACjournalVolNumPages{Atmosphere}{12}{5}{}.
\newblock
\begin{APACrefDOI} \doi{10.3390/atmos12050580} \end{APACrefDOI}
\PrintBackRefs{\CurrentBib}

\bibitem [\protect \citeauthoryear {%
Flesch%
, Wilson%
\BCBL {}\ \BBA {} Yee%
}{%
Flesch%
\ \protect \BOthers {.}}{%
{\protect \APACyear {1995}}%
}]{%
Flesch1995}
\APACinsertmetastar {%
Flesch1995}%
\begin{APACrefauthors}%
Flesch, T\BPBI K.%
, Wilson, J\BPBI D.%
\BCBL {}\ \BBA {} Yee, E.%
\end{APACrefauthors}%
\unskip\
\newblock
\APACrefYearMonthDay{1995}{}{}.
\newblock
\APACrefbtitle {{Backward-time Lagrangian stochastic dispersion models and
  their application to estimate gaseous emissions}} {{Backward-time Lagrangian
  stochastic dispersion models and their application to estimate gaseous
  emissions}}\ (\BVOL~34)\ (\BNUM~6).
\newblock
\begin{APACrefDOI} \doi{10.1175/1520-0450(1995)034<1320:BTLSDM>2.0.CO;2}
  \end{APACrefDOI}
\PrintBackRefs{\CurrentBib}

\bibitem [\protect \citeauthoryear {%
Gavze%
\ \BBA {} Fattal%
}{%
Gavze%
\ \BBA {} Fattal%
}{%
{\protect \APACyear {2018}}%
}]{%
Gavze2018}
\APACinsertmetastar {%
Gavze2018}%
\begin{APACrefauthors}%
Gavze, E.%
\BCBT {}\ \BBA {} Fattal, E.%
\end{APACrefauthors}%
\unskip\
\newblock
\APACrefYearMonthDay{2018}{}{}.
\newblock
{\BBOQ}\APACrefatitle {{A Semi-analytical Model for Short-Range Near-Ground
  Continuous Dispersion}} {{A Semi-analytical Model for Short-Range Near-Ground
  Continuous Dispersion}}.{\BBCQ}
\newblock
\APACjournalVolNumPages{Boundary-Layer Meteorology}{169}{2}{297--326}.
\newblock
\begin{APACrefURL} \url{https://doi.org/10.1007/s10546-018-0363-5}
  \end{APACrefURL}
\newblock
\begin{APACrefDOI} \doi{10.1007/s10546-018-0363-5} \end{APACrefDOI}
\PrintBackRefs{\CurrentBib}

\bibitem [\protect \citeauthoryear {%
Hutchinson%
, Oh%
\BCBL {}\ \BBA {} Chen%
}{%
Hutchinson%
\ \protect \BOthers {.}}{%
{\protect \APACyear {2017}}%
}]{%
Hutchinson2017}
\APACinsertmetastar {%
Hutchinson2017}%
\begin{APACrefauthors}%
Hutchinson, M.%
, Oh, H.%
\BCBL {}\ \BBA {} Chen, W\BPBI H.%
\end{APACrefauthors}%
\unskip\
\newblock
\APACrefYearMonthDay{2017}{}{}.
\newblock
{\BBOQ}\APACrefatitle {{A review of source term estimation methods for
  atmospheric dispersion events using static or mobile sensors}} {{A review of
  source term estimation methods for atmospheric dispersion events using static
  or mobile sensors}}.{\BBCQ}
\newblock
\APACjournalVolNumPages{Information Fusion}{36}{}{130--148}.
\newblock
\begin{APACrefURL} \url{http://dx.doi.org/10.1016/j.inffus.2016.11.010}
  \end{APACrefURL}
\newblock
\begin{APACrefDOI} \doi{10.1016/j.inffus.2016.11.010} \end{APACrefDOI}
\PrintBackRefs{\CurrentBib}

\bibitem [\protect \citeauthoryear {%
Hutchinson%
, Oh%
\BCBL {}\ \BBA {} Chen%
}{%
Hutchinson%
\ \protect \BOthers {.}}{%
{\protect \APACyear {2018}}%
}]{%
Hutchinson2018}
\APACinsertmetastar {%
Hutchinson2018}%
\begin{APACrefauthors}%
Hutchinson, M.%
, Oh, H.%
\BCBL {}\ \BBA {} Chen, W\BPBI H.%
\end{APACrefauthors}%
\unskip\
\newblock
\APACrefYearMonthDay{2018}{}{}.
\newblock
{\BBOQ}\APACrefatitle {{Entrotaxis as a strategy for autonomous search and
  source reconstruction in turbulent conditions}} {{Entrotaxis as a strategy
  for autonomous search and source reconstruction in turbulent
  conditions}}.{\BBCQ}
\newblock
\APACjournalVolNumPages{Information Fusion}{42}{March 2017}{179--189}.
\newblock
\begin{APACrefURL} \url{https://doi.org/10.1016/j.inffus.2017.10.009}
  \end{APACrefURL}
\newblock
\begin{APACrefDOI} \doi{10.1016/j.inffus.2017.10.009} \end{APACrefDOI}
\PrintBackRefs{\CurrentBib}

\bibitem [\protect \citeauthoryear {%
J.~Issartel%
}{%
J.~Issartel%
}{%
{\protect \APACyear {2003}}%
}]{%
Issartel2003}
\APACinsertmetastar {%
Issartel2003}%
\begin{APACrefauthors}%
Issartel, J.%
\end{APACrefauthors}%
\unskip\
\newblock
\APACrefYearMonthDay{2003}{}{}.
\newblock
{\BBOQ}\APACrefatitle {{Rebuilding sources of linear tracers after atmospheric
  concentration measurements}} {{Rebuilding sources of linear tracers after
  atmospheric concentration measurements}}.{\BBCQ}
\newblock
\APACjournalVolNumPages{}{}{1987}{2111--2125}.
\PrintBackRefs{\CurrentBib}

\bibitem [\protect \citeauthoryear {%
J\BPBI P.~Issartel%
}{%
J\BPBI P.~Issartel%
}{%
{\protect \APACyear {2005}}%
}]{%
Issartel2005}
\APACinsertmetastar {%
Issartel2005}%
\begin{APACrefauthors}%
Issartel, J\BPBI P.%
\end{APACrefauthors}%
\unskip\
\newblock
\APACrefYearMonthDay{2005}{}{}.
\newblock
{\BBOQ}\APACrefatitle {{Emergence of a tracer source from air concentration
  measurements, a new strategy for linear assimilation}} {{Emergence of a
  tracer source from air concentration measurements, a new strategy for linear
  assimilation}}.{\BBCQ}
\newblock
\APACjournalVolNumPages{Atmospheric Chemistry and Physics}{5}{1}{249--273}.
\newblock
\begin{APACrefDOI} \doi{10.5194/acp-5-249-2005} \end{APACrefDOI}
\PrintBackRefs{\CurrentBib}

\bibitem [\protect \citeauthoryear {%
A.~Keats%
, Yee%
\BCBL {}\ \BBA {} Lien%
}{%
A.~Keats%
\ \protect \BOthers {.}}{%
{\protect \APACyear {2007}}%
}]{%
Keats2007}
\APACinsertmetastar {%
Keats2007}%
\begin{APACrefauthors}%
Keats, A.%
, Yee, E.%
\BCBL {}\ \BBA {} Lien, F\BPBI S.%
\end{APACrefauthors}%
\unskip\
\newblock
\APACrefYearMonthDay{2007}{}{}.
\newblock
{\BBOQ}\APACrefatitle {{Bayesian inference for source determination with
  applications to a complex urban environment}} {{Bayesian inference for source
  determination with applications to a complex urban environment}}.{\BBCQ}
\newblock
\APACjournalVolNumPages{Atmospheric Environment}{41}{3}{465--479}.
\newblock
\begin{APACrefDOI} \doi{10.1016/j.atmosenv.2006.08.044} \end{APACrefDOI}
\PrintBackRefs{\CurrentBib}

\bibitem [\protect \citeauthoryear {%
A.~Keats%
, Yee%
\BCBL {}\ \BBA {} Lien%
}{%
A.~Keats%
\ \protect \BOthers {.}}{%
{\protect \APACyear {2010}}%
}]{%
Keats2010}
\APACinsertmetastar {%
Keats2010}%
\begin{APACrefauthors}%
Keats, A.%
, Yee, E.%
\BCBL {}\ \BBA {} Lien, F\BPBI S.%
\end{APACrefauthors}%
\unskip\
\newblock
\APACrefYearMonthDay{2010}{}{}.
\newblock
{\BBOQ}\APACrefatitle {{Information-driven receptor placement for contaminant
  source determination}} {{Information-driven receptor placement for
  contaminant source determination}}.{\BBCQ}
\newblock
\APACjournalVolNumPages{Environmental Modelling and
  Software}{25}{9}{1000--1013}.
\newblock
\begin{APACrefURL} \url{http://dx.doi.org/10.1016/j.envsoft.2010.01.006}
  \end{APACrefURL}
\newblock
\begin{APACrefDOI} \doi{10.1016/j.envsoft.2010.01.006} \end{APACrefDOI}
\PrintBackRefs{\CurrentBib}

\bibitem [\protect \citeauthoryear {%
W\BPBI A.~Keats%
}{%
W\BPBI A.~Keats%
}{%
{\protect \APACyear {2009}}%
}]{%
keats2009}
\APACinsertmetastar {%
keats2009}%
\begin{APACrefauthors}%
Keats, W\BPBI A.%
\end{APACrefauthors}%
\unskip\
\newblock
\APACrefYearMonthDay{2009}{}{}.
\newblock
{\BBOQ}\APACrefatitle {{Bayesian inference for source determination in the
  atmospheric environment}} {{Bayesian inference for source determination in
  the atmospheric environment}}.{\BBCQ}
\newblock

\PrintBackRefs{\CurrentBib}

\bibitem [\protect \citeauthoryear {%
Kumar%
, Feiz%
, Singh%
, Ngae%
\BCBL {}\ \BBA {} Turbelin%
}{%
Kumar%
\ \protect \BOthers {.}}{%
{\protect \APACyear {2015}}%
}]{%
Kumar2015}
\APACinsertmetastar {%
Kumar2015}%
\begin{APACrefauthors}%
Kumar, P.%
, Feiz, A\BPBI A.%
, Singh, S\BPBI K.%
, Ngae, P.%
\BCBL {}\ \BBA {} Turbelin, G.%
\end{APACrefauthors}%
\unskip\
\newblock
\APACrefYearMonthDay{2015}{}{}.
\newblock
{\BBOQ}\APACrefatitle {{Reconstruction of an atmospheric tracer source in an
  urban-like environment}} {{Reconstruction of an atmospheric tracer source in
  an urban-like environment}}.{\BBCQ}
\newblock
\APACjournalVolNumPages{Journal of Geophysical
  Research}{120}{24}{12,589--12,604}.
\newblock
\begin{APACrefDOI} \doi{10.1002/2015JD024110} \end{APACrefDOI}
\PrintBackRefs{\CurrentBib}

\bibitem [\protect \citeauthoryear {%
Lindley%
}{%
Lindley%
}{%
{\protect \APACyear {1956}}%
}]{%
Lindley1956}
\APACinsertmetastar {%
Lindley1956}%
\begin{APACrefauthors}%
Lindley, D\BPBI V.%
\end{APACrefauthors}%
\unskip\
\newblock
\APACrefYearMonthDay{1956}{}{}.
\newblock
{\BBOQ}\APACrefatitle {{On a Measure of the Information Provided by an
  Experiment}} {{On a Measure of the Information Provided by an
  Experiment}}.{\BBCQ}
\newblock
\APACjournalVolNumPages{The Annals of Mathematical
  Statistics}{27}{4}{986--1005}.
\newblock
\begin{APACrefDOI} \doi{10.1214/aoms/1177728069} \end{APACrefDOI}
\PrintBackRefs{\CurrentBib}

\bibitem [\protect \citeauthoryear {%
Loredo%
}{%
Loredo%
}{%
{\protect \APACyear {2004}}%
}]{%
Loredo2004}
\APACinsertmetastar {%
Loredo2004}%
\begin{APACrefauthors}%
Loredo, T\BPBI J.%
\end{APACrefauthors}%
\unskip\
\newblock
\APACrefYearMonthDay{2004}{}{}.
\newblock
{\BBOQ}\APACrefatitle {{Bayesian Adaptive Exploration}} {{Bayesian Adaptive
  Exploration}}.{\BBCQ}
\newblock
\APACjournalVolNumPages{}{}{}{330--346}.
\newblock
\begin{APACrefDOI} \doi{10.1063/1.1751377} \end{APACrefDOI}
\PrintBackRefs{\CurrentBib}

\bibitem [\protect \citeauthoryear {%
Menke%
}{%
Menke%
}{%
{\protect \APACyear {2012}}%
}]{%
Menke}
\APACinsertmetastar {%
Menke}%
\begin{APACrefauthors}%
Menke.%
\end{APACrefauthors}%
\unskip\
\newblock
\APACrefYear{2012}.
\newblock
\APACrefbtitle {{Discrete Inverse problem in geophysics}} {{Discrete Inverse
  problem in geophysics}}.
\PrintBackRefs{\CurrentBib}

\bibitem [\protect \citeauthoryear {%
Pope%
}{%
Pope%
}{%
{\protect \APACyear {2000}}%
}]{%
pope_2000}
\APACinsertmetastar {%
pope_2000}%
\begin{APACrefauthors}%
Pope, S\BPBI B.%
\end{APACrefauthors}%
\unskip\
\newblock
\APACrefYear{2000}.
\newblock
\APACrefbtitle {{Turbulent Flows}} {{Turbulent Flows}}.
\newblock
\APACaddressPublisher{}{Cambridge University Press}.
\newblock
\begin{APACrefDOI} \doi{10.1017/CBO9780511840531} \end{APACrefDOI}
\PrintBackRefs{\CurrentBib}

\bibitem [\protect \citeauthoryear {%
Ristic%
, Skvortsov%
\BCBL {}\ \BBA {} Gunatilaka%
}{%
Ristic%
\ \protect \BOthers {.}}{%
{\protect \APACyear {2016}}%
}]{%
Ristic2016a}
\APACinsertmetastar {%
Ristic2016a}%
\begin{APACrefauthors}%
Ristic, B.%
, Skvortsov, A.%
\BCBL {}\ \BBA {} Gunatilaka, A.%
\end{APACrefauthors}%
\unskip\
\newblock
\APACrefYearMonthDay{2016}{}{}.
\newblock
{\BBOQ}\APACrefatitle {{A study of cognitive strategies for an autonomous
  search}} {{A study of cognitive strategies for an autonomous search}}.{\BBCQ}
\newblock
\APACjournalVolNumPages{Information Fusion}{28}{}{1--9}.
\newblock
\begin{APACrefURL} \url{http://dx.doi.org/10.1016/j.inffus.2015.06.008}
  \end{APACrefURL}
\newblock
\begin{APACrefDOI} \doi{10.1016/j.inffus.2015.06.008} \end{APACrefDOI}
\PrintBackRefs{\CurrentBib}

\bibitem [\protect \citeauthoryear {%
Rudd%
, Robins%
, Lepley%
\BCBL {}\ \BBA {} Belcher%
}{%
Rudd%
\ \protect \BOthers {.}}{%
{\protect \APACyear {2012}}%
}]{%
Rudd2012}
\APACinsertmetastar {%
Rudd2012}%
\begin{APACrefauthors}%
Rudd, A\BPBI C.%
, Robins, A\BPBI G.%
, Lepley, J\BPBI J.%
\BCBL {}\ \BBA {} Belcher, S\BPBI E.%
\end{APACrefauthors}%
\unskip\
\newblock
\APACrefYearMonthDay{2012}{}{}.
\newblock
{\BBOQ}\APACrefatitle {{An Inverse Method for Determining Source
  Characteristics for Emergency Response Applications}} {{An Inverse Method for
  Determining Source Characteristics for Emergency Response
  Applications}}.{\BBCQ}
\newblock
\APACjournalVolNumPages{Boundary-Layer Meteorology}{144}{1}{1--20}.
\newblock
\begin{APACrefDOI} \doi{10.1007/s10546-012-9712-y} \end{APACrefDOI}
\PrintBackRefs{\CurrentBib}

\bibitem [\protect \citeauthoryear {%
Sawford%
}{%
Sawford%
}{%
{\protect \APACyear {1985}}%
}]{%
Sawford1985}
\APACinsertmetastar {%
Sawford1985}%
\begin{APACrefauthors}%
Sawford, B\BPBI L.%
\end{APACrefauthors}%
\unskip\
\newblock
\APACrefYearMonthDay{1985}{}{}.
\newblock
\APACrefbtitle {{Lagrangian statistical simulation of concentration mean and
  fluctuation fields.}} {{Lagrangian statistical simulation of concentration
  mean and fluctuation fields.}}\ (\BVOL~24)\ (\BNUM~11).
\newblock
\begin{APACrefDOI} \doi{10.1175/1520-0450(1985)024<1152:LSSOCM>2.0.CO;2}
  \end{APACrefDOI}
\PrintBackRefs{\CurrentBib}

\bibitem [\protect \citeauthoryear {%
Sebastiani%
\ \BBA {} Wynn%
}{%
Sebastiani%
\ \BBA {} Wynn%
}{%
{\protect \APACyear {2000}}%
}]{%
Sebastiani2000}
\APACinsertmetastar {%
Sebastiani2000}%
\begin{APACrefauthors}%
Sebastiani, P.%
\BCBT {}\ \BBA {} Wynn, H\BPBI P.%
\end{APACrefauthors}%
\unskip\
\newblock
\APACrefYearMonthDay{2000}{}{}.
\newblock
{\BBOQ}\APACrefatitle {{Maximum entropy sampling and optimal Bayesian
  experimental design}} {{Maximum entropy sampling and optimal Bayesian
  experimental design}}.{\BBCQ}
\newblock
\APACjournalVolNumPages{Journal of the Royal Statistical Society. Series B:
  Statistical Methodology}{62}{1}{145--157}.
\newblock
\begin{APACrefDOI} \doi{10.1111/1467-9868.00225} \end{APACrefDOI}
\PrintBackRefs{\CurrentBib}

\bibitem [\protect \citeauthoryear {%
Thomson%
}{%
Thomson%
}{%
{\protect \APACyear {1987}}%
}]{%
Thomson1987}
\APACinsertmetastar {%
Thomson1987}%
\begin{APACrefauthors}%
Thomson, D\BPBI J.%
\end{APACrefauthors}%
\unskip\
\newblock
\APACrefYearMonthDay{1987}{}{}.
\newblock
{\BBOQ}\APACrefatitle {{Criteria for the selection of stochastic models of
  particle trajectories in turbulent flows}} {{Criteria for the selection of
  stochastic models of particle trajectories in turbulent flows}}.{\BBCQ}
\newblock
\APACjournalVolNumPages{Journal of Fluid Mechanics}{180}{2}{529--556}.
\newblock
\begin{APACrefDOI} \doi{10.1017/S0022112087001940} \end{APACrefDOI}
\PrintBackRefs{\CurrentBib}

\bibitem [\protect \citeauthoryear {%
Vergassola%
, Villermaux%
\BCBL {}\ \BBA {} Shraiman%
}{%
Vergassola%
\ \protect \BOthers {.}}{%
{\protect \APACyear {2007}}%
}]{%
Vergassola2007}
\APACinsertmetastar {%
Vergassola2007}%
\begin{APACrefauthors}%
Vergassola, M.%
, Villermaux, E.%
\BCBL {}\ \BBA {} Shraiman, B\BPBI I.%
\end{APACrefauthors}%
\unskip\
\newblock
\APACrefYearMonthDay{2007}{}{}.
\newblock
{\BBOQ}\APACrefatitle {{'Infotaxis' as a strategy for searching without
  gradients}} {{'Infotaxis' as a strategy for searching without
  gradients}}.{\BBCQ}
\newblock
\APACjournalVolNumPages{Nature}{445}{7126}{406--409}.
\newblock
\begin{APACrefDOI} \doi{10.1038/nature05464} \end{APACrefDOI}
\PrintBackRefs{\CurrentBib}

\bibitem [\protect \citeauthoryear {%
Yee%
}{%
Yee%
}{%
{\protect \APACyear {2004}}%
}]{%
Yee2004}
\APACinsertmetastar {%
Yee2004}%
\begin{APACrefauthors}%
Yee, E.%
\end{APACrefauthors}%
\unskip\
\newblock
\APACrefYearMonthDay{2004}{}{}.
\newblock
{\BBOQ}\APACrefatitle {{DISPERSING THROUGH A REGULAR ARRAY OF OBSTACLES}}
  {{DISPERSING THROUGH A REGULAR ARRAY OF OBSTACLES}}.{\BBCQ}
\newblock
\APACjournalVolNumPages{}{}{}{363--415}.
\PrintBackRefs{\CurrentBib}

\bibitem [\protect \citeauthoryear {%
Yee%
}{%
Yee%
}{%
{\protect \APACyear {2008}}%
}]{%
Yee2008}
\APACinsertmetastar {%
Yee2008}%
\begin{APACrefauthors}%
Yee, E.%
\end{APACrefauthors}%
\unskip\
\newblock
\APACrefYearMonthDay{2008}{}{}.
\newblock
{\BBOQ}\APACrefatitle {{Theory for reconstruction of an unknown number of
  contaminant sources using probabilistic inference}} {{Theory for
  reconstruction of an unknown number of contaminant sources using
  probabilistic inference}}.{\BBCQ}
\newblock
\APACjournalVolNumPages{Boundary-Layer Meteorology}{127}{3}{359--394}.
\newblock
\begin{APACrefDOI} \doi{10.1007/s10546-008-9270-5} \end{APACrefDOI}
\PrintBackRefs{\CurrentBib}

\bibitem [\protect \citeauthoryear {%
Yee%
}{%
Yee%
}{%
{\protect \APACyear {2012}}%
}]{%
Yee2012}
\APACinsertmetastar {%
Yee2012}%
\begin{APACrefauthors}%
Yee, E.%
\end{APACrefauthors}%
\unskip\
\newblock
\APACrefYearMonthDay{2012}{}{}.
\newblock
{\BBOQ}\APACrefatitle {{Inverse Dispersion for an Unknown Number of Sources:
  Model Selection and Uncertainty Analysis}} {{Inverse Dispersion for an
  Unknown Number of Sources: Model Selection and Uncertainty Analysis}}.{\BBCQ}
\newblock
\APACjournalVolNumPages{ISRN Applied Mathematics}{2012}{}{1--20}.
\newblock
\begin{APACrefDOI} \doi{10.5402/2012/465320} \end{APACrefDOI}
\PrintBackRefs{\CurrentBib}

\end{thebibliography}

%
%
%
%
%

\end{document}